\documentclass[%
 reprint,
superscriptaddress,
 amsmath,amssymb,
 aps,
pra,
floatfix,
]{revtex4-1}

\usepackage{graphicx}
\usepackage{dcolumn}
\usepackage{bm}

\usepackage{siunitx}

\newcommand{\eg}{\emph{e.g.}}
\newcommand{\ie}{\emph{i.e.}}

\newcommand{\Er}{\ensuremath{E_{\mathrm{rec}}}}
\newcommand{\pr}{\ensuremath{p_{\mathrm{rec}}}}

\begin{document}


\title{Bimodal momentum distribution of laser-cooled atoms in optical lattices}

\author{Claude M. Dion}
 \email{claude.dion@umu.se}
 \affiliation{%
 Department of Physics, Ume{\aa} University, SE-901\,87 Ume{\aa}, Sweden
}%
 
\author{Svante Jonsell}%
\affiliation{%
 Department of Physics, Stockholm University, SE-106\,91 Stockholm, Sweden\\
}%

\author{Anders Kastberg}
\email{Anders.KASTBERG@unice.fr}
\affiliation{
 Universit\'e de Nice Sophia Antipolis, CNRS, Laboratoire de Physique de la Mati\`ere Condens\'ee, UMR 7336, Parc Valrose, 06100 Nice, France
}%

\author{Peder Sj\"olund}
 \affiliation{%
 Department of Physics, Ume{\aa} University, SE-901\,87 Ume{\aa}, Sweden
}%

\date{\today}

\begin{abstract}
We study, numerically and experimentally, the momentum distribution of atoms cooled in optical lattices.  Using semi-classical simulations, we show that this distribution is bimodal, made up of a central feature corresponding to ``cold'', trapped atoms, with tails of ``hot'', untrapped atoms, and that this holds true also for very shallow potentials.  Careful analysis of the distribution of high-momentum untrapped atoms, both from simulations and experiments, shows that the tails of the distribution does not follow a normal law, hinting at a power-law distribution and non-ergodic behavior.  We also revisit the phenomenon of \emph{d\'ecrochage}, the potential depth below which the temperature of the atoms starts increasing.
\end{abstract}

\maketitle


\section{\label{sec:intro}Introduction}

Laser cooling in multilevel atomic systems has been a well used technology for the last 25 years; good reviews of its development are given in Refs.~\cite{Chu_RMP_1998,Cohen-Tannoudji_RMP_1998,Phillips_RMP_1998,Fallani_EPL_2015}. A framework for the theory was developed early on (see, \eg, \cite{Dalibard_JOSAB_1989,Castin_Elba_1991} and additional references in~\cite{Chu_RMP_1998,Cohen-Tannoudji_RMP_1998,Phillips_RMP_1998,Fallani_EPL_2015}), involving coherences and optical pumping between levels in the ground state multiplets. Although these models have provided great insights, and reproduce many important experimental findings, they still fail to explain some experimental results, where finer details of the cooling mechanism and the resulting velocity distributions have been studied.

A laser-cooled atomic sample typically has a steady-state velocity distribution to which a Gaussian function provides a very good fit. This has made it possible to assign ``kinetic temperatures'' to the ensembles, even though a strict thermal equilibrium does not exist.

When the laser cooling intensity is very low, small but significant deviations from Gaussian distributions have been observed experimentally~\cite{Jersblad_PRA_2004,Douglas_PRL_2006,Hamilton_PRA_2014} and predicted theoretically~\cite{Lutz_PRA_2003,Lutz_NP_2013}. The experimental observations have been made with dissipative optical lattices~\cite{Grynberg_PR_2001}, or a corresponding laser cooling configuration, and the physical explanation for the observed non-Gaussian distributions has been a contentious point.

\subsection{\label{sec:interest}The interest of laser cooling}


The lack of a complete understanding of the cooling mechanisms has not been a major practical problem in cold atom physics, since laser cooling is a technology that does work excellently as a tool. It has been the key to the advent of Bose-Einstein condensation in dilute atomic gases, and is a major ingredient, for example, in experiments with frequency standards, quantum information, and fundamental metrology.

Recently, the interest in fundamental laser cooling has been revived. This is to a large extent due to practical reasons, as scientists are turning towards systems that are increasingly more difficult to laser cool than the alkali atoms that have been the workhorses in most laser cooling related research hitherto (Na, Rb and Cs). As the complexity increases, so does the need for an understanding of how to adapt the technology in order to achieve the desired cooling. One example is molecular laser cooling~%
\cite{Shuman_Nature_2010,Comparat_PRA_2014,Zhelyazkova_PRA_2014,Hamamda_JPB_2015,Yzombard_PRL_2015},
where the presence of vibrational and rotational degrees of freedom typically result in extremely complex energy level diagrams. 

In the case of  atoms, more complex systems than the staple elements are being studied more closely. Such examples are Li and K (see, \eg, \cite{Landini_PRA_2011,RioFernandes_EPL_2012,Hamilton_PRA_2014,Salomon_PRA_2014,Burchianti_PRA_2014,Sievers_PRA_2015}), which are alkalis as well, but that have hyperfine structure splittings that complicate the cooling process. Increased knowledge of the laser cooling process may also be required when atoms are cooled in a setting, or geometry, that significantly changes the conditions, or in the cooling of other types of systems (see, \eg, \cite{Vetsch_PRL_2010,Volchkov_NJP_2013,Reiserer_PRL_2013,Sparkes_PRA_2015,Yin_JPB_2015,Lepers_PRA_2016}). With these new challenges to laser cooling, the motivation for deepening the understanding of the cooling process increases, as does the need for honing theoretical and numerical tools for its analysis.

\subsection{\label{sec:invest_distr}Investigating the velocity distribution}

In this work, we study the velocity distribution of atoms laser-cooled in a shallow optical lattice in detail, both experimentally and theoretically. In particular, we address the issue of whether, for a shallow optical lattice, the entire atomic population can be adequately described by a single distribution function, and if spatial averaging can be applied, or if the fraction of the atoms that are localized at potential minima has to be accounted for. Thereby, we seek to clarify the impact of localization on observed non-Gaussian momentum profiles. These issues were addressed by some of us already in Ref.~\cite{Jersblad_PRA_2004}, with conclusions that have been supported by others~\cite{Greenberg_OE_2011,Greenberg_EPL_2012,Hamilton_PRA_2014}. However, controversies have followed concerning the interpretation of velocity distributions~\cite{Douglas_PRL_2006,Bakar_PRE_2009,Lutz_NP_2013,Holz_EPL_2015,Dechant_PRL_2015}, which have motivated us to revisit the question.

\section{\label{sec:theory}Theory}

\subsection{\label{sec:sisyphus}The standard model of laser cooling -- Sisyphus cooling}

When sub-Doppler cooling was first discovered experimentally~\cite{Lett_PRL_1988}, the theoretical explanation that followed shortly afterwards was based on the concepts of polarization gradients and atomic state degeneracy. Spatially-dependent optical pumping induces slow time scales and a time lag between the internal and external evolution of the atoms. The model is often referred to as Sisyphus cooling~\cite{Dalibard_JOSAB_1989}.

Sisyphus cooling was explained qualitatively by the seminal model of Dalibard and Cohen-Tannoudji~\cite{Dalibard_JOSAB_1989}. In its most simple form, this model assumes that atoms are moving as classical particles through a one-dimensional modulated optical potential. The latter emanates from two red-detuned laser beams with orthogonal linear polarization --- the so-called lin$\perp$lin configuration. 

The key to the cooling mechanism is the internal structure of the atom, coupled to the spatially-alternating polarization of the light. For an atom with a ground-state angular momentum $J_\mathrm{g} = 1/2$, there exist two magnetic sublevels $M_\mathrm{g} = \pm 1/2$. The degeneracy of these two states is broken by the AC Stark shift arising from the interaction with the laser-cooling light, in such a way that two sinusoidal potentials, phase-shifted by half a period, are created. Absorption of laser photons followed by spontaneous emission then leads to optical pumping between the two magnetic states. For a correctly chosen detuning of the light, the probability for optical pumping is at its largest at the peak of the a potential, while it is lowest at its bottom. Since the peak of the potential of, \eg, the $M_\mathrm{g} = 1/2$ state coincides spatially with the bottom of the $M_\mathrm{g} = -1/2$ potential, this will have the consequence that the atom on average spends more time climbing the peaks of the potential than it spends falling down towards its valleys. After averaging over the two internal states, and over a spatial period of the lattice, this therefore leads to an effective friction force. In addition, the fluctuations induced by the randomness of the optical pumping process also lead to diffusion. The balance between friction and momentum diffusion determines the steady-state temperature of the atoms.

This model, albeit simple, appears to capture the essence of the physical mechanism behind Sisyphus cooling. It has been vindicated by good qualitative agreements with more advanced theoretical simulations, as well as with experiments (see, \eg, \cite{Lett_JOSAB_1989}). For instance, the linear scaling between temperature and potential depth is correctly predicted by this model~\cite{Salomon_EPL_1990}. 

However, for a detailed agreement between model and experiments, there are a number of complications which need to be considered, and some unresolved problems. This includes the three-dimensionality of the optical potential, the more complicated level structure of real atoms, the rate of cooling, quantum effects, and taking the full spatial modulation of the atomic densities into account. In the 1990s, there was a considerable effort to enhance the understanding of the cooling mechanisms involved (see, \eg, \cite{Castin_EPL_1991,Nienhuis_PRA_1991,Gerz_EPL_1993,Chen_PRA_1993,Javanainen_JPB_1994,Castin_PRA_1994,Kuppens_PRA_1998}, and other references within those articles). In these works, different theoretical approaches (semi-classical as well as fully quantum mechanical ones) were compared with detailed experiments and important insights were gained. To our knowledge, an extensive review of all hitherto known aspects of polarization-gradient cooling is lacking, and such a treatise is also beyond the scope of the present work. This study is focused on the issue of the shape of the steady-state velocity distribution that arises from the cooling.

The constant friction and diffusion coefficients obtained by the spatial averaging procedure in Ref.~\cite{Dalibard_JOSAB_1989} entails a perfectly Gaussian momentum distribution of the atoms. While on the whole this profile agrees remarkably well with the majority of the experimental findings, it cannot explain the small deviations in the wings of the momentum distribution found experimentally in, \eg, Refs.~\cite{Jersblad_PRA_2004,Douglas_PRL_2006,Hamilton_PRA_2014}. This is by no means surprising, considering the many simplifications of the theory required to derive the perfectly Gaussian profile, as summarized below. Deviations from Gaussian velocity distributions are also noted in some of the works referenced in the preceding paragraph, as well as Ref.~\cite{Castin_thesis_1992}.

\subsubsection{\label{sec:Sisyph_Gauss}Gaussian velocity distribution}
Several works describe how the one-dimensional Sisyphus cooling model leads to a Gaussian momentum distribution, see for example Refs.~\cite{Dalibard_JOSAB_1989,Castin_Elba_1991,Hodapp_APB_1995,Marksteiner_PRA_1996,Jersblad_PRA_2004}. We present here only a brief outline.

The cooling is expressed as a friction force, 
\begin{equation}
F(p)=-\frac{\alpha p}{m} .
\end{equation}
This is only true within a narrow velocity range, called the velocity capture range (or momentum capture range, $p_\mathrm{c}$), and the assumption is made that the entire sample is within this domain.

The cooling is counterbalanced by a momentum diffusion, $D_p$, which is time-averaged and taken as independent of velocity. $D_p$  has two main contributions,
\begin{equation}
D_p(p) = D_p^{(\mathrm{ph})} + D_p^{(\mathrm{pot})} ,
\end{equation}
with $D_p^{(\mathrm{ph})}$ arising from the stochastic nature of light scattering and $D_p^{(\mathrm{pot})}$ originating from fluctuations in the instantaneous potential felt by an atom. The competition between cooling and heating can then in turn be described by a Fokker-Planck equation,
\begin{equation}
\frac{\partial W(p,t)}{\partial t} = - \frac{\partial}{\partial p}\left[F(p) W(p,t) \right]+ \frac{\partial}{\partial p} \left[ D(p) \, \frac{\partial W(p,t) }{\partial p}  \right] .
\label{eq:fokker-planck}
\end{equation}
In steady-state, for a momentum-independent diffusion $D_p$, the solution of Eq.~(\ref{eq:fokker-planck}) leads to a Gaussian distribution,
\begin{equation}
\left\langle W(p) \right\rangle_t = W_0 \, \exp\left( -\frac{\alpha p^2}{2m^3D_p}\right) .
\end{equation}

Below a certain laser intensity, the Sisyphus cooling becomes too weak to retain a normalizable momentum distribution. This means that for weaker laser intensities, the linear scaling of the temperature is broken, and instead the temperature increases rapidly with shallower optical potentials. This has been verified in many experiments, \eg, Refs.~\cite{Salomon_EPL_1990,Jersblad_PRA_2000}. 

The phenomenon that, for intensities below a critical one, the measured temperature quickly increases has often been referred to as \emph{d\'ecrochage} (see, \eg, \cite{Hodapp_APB_1995})\footnote{The English translation of the French word ``\emph{d\'ecrochage}'' is ``unhooking''}. In the early literature, this effect was frequently taken as a consequence of the velocity capture range becoming too narrow to catch the entire Boltzmann distribution and thus the sample, the \emph{optical molasses}, would disintegrate. In this article, we will \emph{a priori} use the term with its phenomenological definition, and we will discuss its causes in Sec.~\ref{sec:discussion}.

\subsection{\label{sec:NG_tails}Non-Gaussian velocity tails}

A simplification made in Ref.~\cite{Dalibard_JOSAB_1989} is the assumption that the total density profile of the atoms (summing both magnetic states) is spatially uniform, while the spatial dependence of the two sublevels simply mirror the spatial dependence of the pumping rates between the two states. A first step towards a more complete theory, while retaining most of the conceptual simplicity of the model of Dalibard and Cohen-Tannoudji, is to include also the effect of the motion in the potentials when determining the spatial dependence of the density profiles in the different potentials (while still assuming that the total population has no spatial modulation)~\cite{Castin_Elba_1991}. In doing this, a momentum dependence is introduced into the populations, and hence into the friction and diffusion coefficients. When this momentum dependence is included, the wings of the atomic momentum distribution change from a Gaussian to a power-law form~\cite{Castin_Elba_1991,Lutz_PRA_2003}.

Another consequence is more insight into the existence of a lower limit for the intensity, for which the equilibrium temperature is minimized. This \emph{d\'ecrochage} phenomenon now becomes more directly related to the modulation depth of the optical potentials. As can be expected, the non-Gaussian features observed in the momentum distributions are especially prominent for potential depths close to or below this critical point.


Taking into account momenta beyond $p_\mathrm{c}$, the expressions for friction and momentum diffusion have to be replaced by
\begin{align}
F (p) &= - \frac { \alpha p } { m \left( 1 + \left[\frac{p}{p_\mathrm{c}}\right]^2\right)} \\ 
\intertext{and}
D_p(p) &= D_p^{(\mathrm{ph})} + \frac{D_p^{(\mathrm{pot})}}{1 + \left(\frac{p}{p_\mathrm{c}}\right)^2} .
\end{align}

\subsubsection{\label{sec:Tsallis}Tsallis distribution}

In Ref.~\cite{Lutz_PRA_2003}, Lutz showed that the semi-classical model of Sisyphus cooling presented in Sec.~\ref{sec:sisyphus} leads to a steady-state Wigner function for the momentum distribution of the atoms [see Eq.~(\ref{eq:fokker-planck})] given by
\begin{equation}
W_q(p) = Z_q^{-1} \left[ 1 - \beta (1-q) p^2 \right]^{1/(1-q)} ,
\label{eq:lutz}
\end{equation}
which is in the form of a Tsallis function~\cite{Tsallis_JSP_1988}. The factor $Z_q^{-1}$ corresponds to an amplitude, and the parameters $\beta$ and $q$ can be derived from the friction and the diffusion coefficients as
\begin{align}
q &= 1 + \frac{2 m^3 D_p^{(\mathrm{ph})}}{\alpha p_\mathrm{c}^2} \\
\intertext{and}
\beta &= \frac{\alpha}{2m\left( D_p^{(\mathrm{ph})} + D_p^{(\mathrm{pot})} \right)}.
\end{align}
We stress that this momentum distribution is obtained when the possible trapping and localization of atoms in optical lattice sites has been neglected.

While a Gaussian is recovered from Eq.~(\ref{eq:lutz}) when $q \rightarrow 0$, it leads to the possibility of non-Gaussian velocity distributions, especially in the high-velocity part of those distributions (the ``tails'') where trapping is no longer relevant. It has also been shown to lead to anomalous diffusion \cite{Marksteiner_PRA_1996,Katori_PRL_1997} in the optical lattice. We note that even in the case where some atoms are trapped, Eq.~(\ref{eq:lutz}) may still be a good description for a part of the atomic population that remains untrapped.

\subsection{\label{sec:localization}Localization at lattice sites}

The model using momentum-dependent friction and diffusion in Ref.~\cite{Castin_Elba_1991} goes some way to include the effects of the modulation of the potential on the atomic populations. However, it still assumes that the total atomic population is spatially uniform, and employs spatial averaging over a period of the lattice. In the limit of atomic energies (kinetic+potential) smaller than the depth of the lattice, the atoms will localize near the bottom of the potential wells. As the regions close to the peaks of the lattice will be inaccessible to these atoms (independently of their internal state), it is clear that the assumption of a spatially uniform total atomic distribution will not hold. This localization effect has been theoretically and experimentally verified in deep optical lattices \cite{Westbrook_PRL_1990,Jessen_PRL_1992,Verkerk_PRL_1992,Marte_PRL_1993, Hemmerich_PRL_1993,Grynberg_PRL_1993,Hemmerich_EPL_1993}. It is, however, less clear if localization plays an important role in lattices where the momentum profile of the atoms have prominent non-Gaussian wings, \ie, at or below \emph{d\'erochage}. 

With the presence of the optical lattice light, there will always be heating present, with the possibility for a trapped atom to become untrapped. This untrapped atom will in turn be exposed to laser cooling. Thus, we assume that at any given time, a subset of the atomic population will be moving across the lattice, whereas another portion of atoms will be localized. Moreover, there will be transfers between these two populations and a corresponding steady state (provided no atom can escape from the optical lattice). For deep optical lattices, the portion of untrapped atoms will be very small --- typically too small to measure. For very shallow lattices however --- close to \emph{d\'ecrochage} --- there will be significant portion of both classes of atoms, and thus a snapshot of the velocity distribution should show a bimodal distribution. Under the assumption above, the untrapped atoms ought to follow a power-law distribution, as in Eq.~(\ref{eq:lutz}), whereas the trapped portion should be fitted separately, for example to a truncated Gaussian.

\section{\label{sec:sc_sim}Semi-classical simulations}

\subsection{\label{sec:num_met}Numerical methods}

We calculate the steady state of atoms in a one-dimensional optical lattice using the semi-classical method described in Refs.~\cite{Castin_Elba_1991,Petsas_EPJD_1999,Jonsell_EPJD_2006,Svensson_EPJD_2008}. While the lattice is 1D and only motion along the lattice axis is considered, photons can be spontaneously emitted in any direction in 3D.  The position and momentum are treated as classical variables, described by a Wigner distribution, while a quantum representation is used for the internal state of the atom.   We consider two cases for the latter: either a $J_{\mathrm{g}} = 1/2 \leftrightarrow J_{\mathrm{e}} = 3/2$ transition, the minimal degeneracy exhibiting Sisyphus cooling, or the $F_{\mathrm{g}} = 4 \leftrightarrow F_{\mathrm{e}} = 5$ transition corresponding to cesium cooled on the D2 line, including the presence of the $F_{\mathrm{e}} = 4$ state~\cite{Svensson_EPJD_2008}.  In the first case, the laser does not couple the two ground states (an atom only shifts between them through spontaneous emission), such that an atom is found in either of the $\pm 1/2$ substates at any given time.  In the other case, the atoms end up in superpositions of either even or odd $M_F$ substates (adiabatic potentials).

The simulations depend on two parameters: the detuning $\Delta$ of the laser with respect to the atomic transition and $\Delta' \equiv \Delta s_0/2$, with $s_0$ the saturation parameter~\footnote{Note that we calculate the staturation based on Rabi frequency of the total laser field. This is the same convention as was used, \eg, in Refs.~\cite{Petsas_EPJD_1999,Jonsell_EPJD_2006,Svensson_EPJD_2008}.  Some other authors have used  the Rabi frequency based on the laser irradiance \emph{per beam} which, for the one-dimensional lin$\perp$lin configuration considered here, is half the total irradiance.}.  The former is usually expressed in units of the natural linewidth of the excited state, $\Gamma$, while the latter is directly proportional to the amplitude of the optical lattice potential. The potential depth is given by 
\begin{equation}
U = A \hbar \left| \Delta' \right|,
\label{eq:depth}
\end{equation}
where $A = 2/3$ for the $1/2 \leftrightarrow 3/2$ transition and $A = 4/9$ (based on the lowest adiabatic potential) for the $4 \leftrightarrow 5$ transition.  Energies are conveniently expressed in terms of the recoil energy,
\begin{equation}
\Er \equiv \frac{\pr^2}{2m},
\end{equation}
\ie, the kinetic energy gained by the atom when spontaneously emitting a photon, where $\pr \equiv \hbar k$, with $k$ the wave vector of the optical lattice laser.

Unless otherwise noted, the results are obtained for 200000 and 100000 independent atoms for the $1/2 \leftrightarrow 3/2$ and $4 \leftrightarrow 5$ transitions, respectively.  Momentum distributions along the axis of the optical lattice are obtained by accumulating the final momentum of atoms into bins of width $\pr$.

To determine if an atom is trapped or not in one of the potential wells of the lattice, we need to compare its total (potential+kinetic) energy with the depth of these potential wells, Eq.~(\ref{eq:depth}).  This is straightforward for the $1/2 \leftrightarrow 3/2$ case, where the amplitude $U$ of the potential is the same for both internal states. For the $4 \leftrightarrow 5$ transition, the light shift varies with the $M_F$ substate~\cite{Grynberg_PR_2001} and an atom is found in a superposition of $M_F$ states, with optical pumping pushing atoms towards the extreme $M_F = \pm F$ states~\cite{Kastler_JPR_1950}.  Moreover, we find that the adiabatic potentials~\cite{Grynberg_PR_2001} better represent the interaction of the atom with the laser field.  Therefore, for the $4 \leftrightarrow 5$ transition, we define as trapped atoms that have an energy lower than the barrier height in the lowest adiabatic potential.  This results in a slight overestimation of the number of trapped atoms in this case, as some atoms that are not in the lowest adiabatic state can be counted as trapped even though they have enough energy to escape to a neighboring well in another adiabatic state. 

We can calculate the maximum momentum $p_{\mathrm{trap}}$ an atom can have and still be trapped as
\begin{equation}
\frac{p_{\mathrm{trap}}}{\pr} \equiv \left(  \frac{U}{\Er} \right)^{1/2} = \left( \frac{A \hbar \left| \Delta \right|'}{\Er} \right)^{1/2}.
\label{eq:ptrap}
\end{equation}
Note that an atom with a momentum $0\leq p \leq p_{\mathrm{trap}}$ can either be trapped or untrapped, depending on the amount of potential energy it has at its current position and state.

\subsection{\label{sec:num_res}Numerical results}

\subsubsection{Trapped vs. untrapped atoms}

\begin{figure}
	\centerline{\includegraphics[width=\columnwidth]{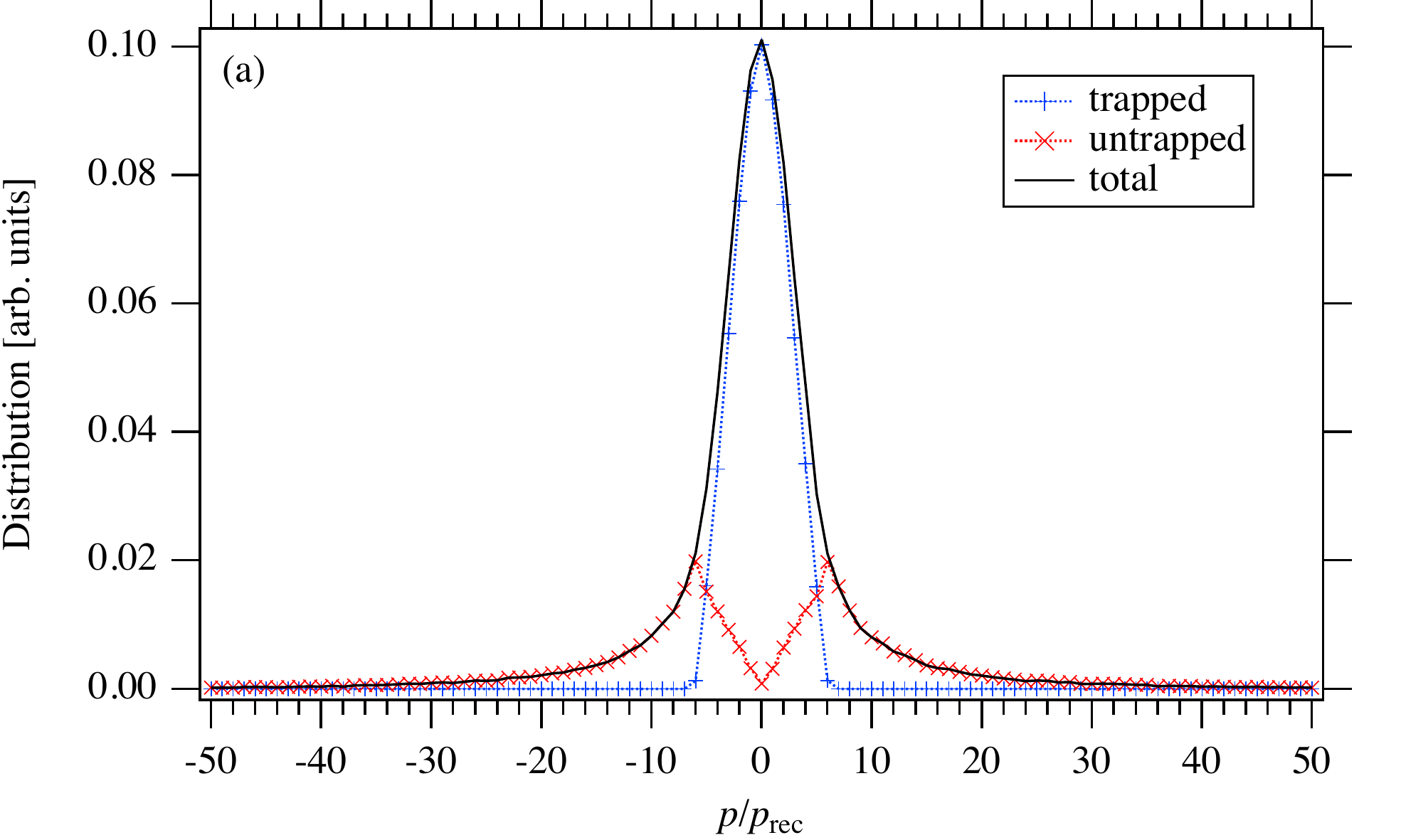}}
    \centerline{\includegraphics[width=\columnwidth]{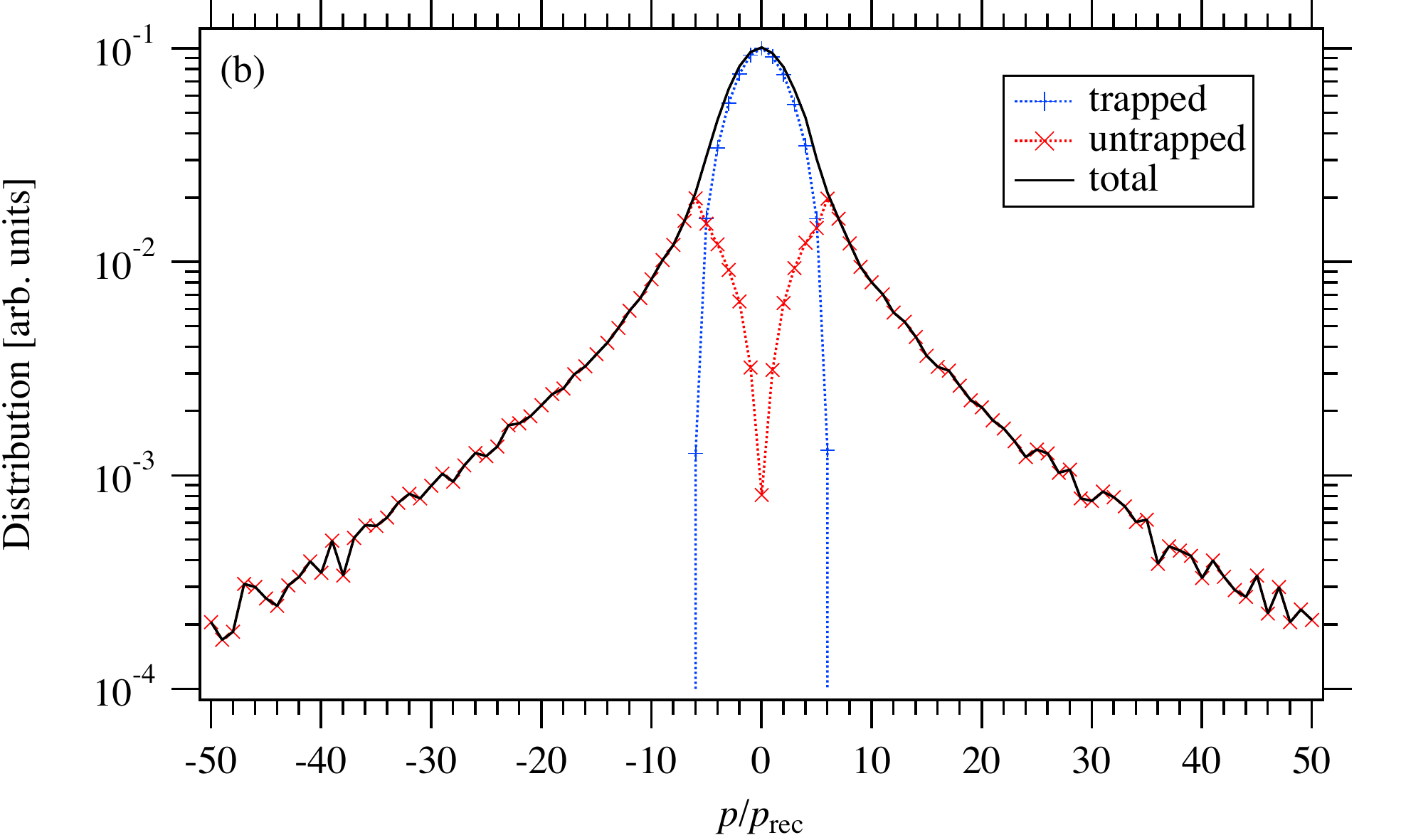}}
    \centerline{\includegraphics[width=\columnwidth]{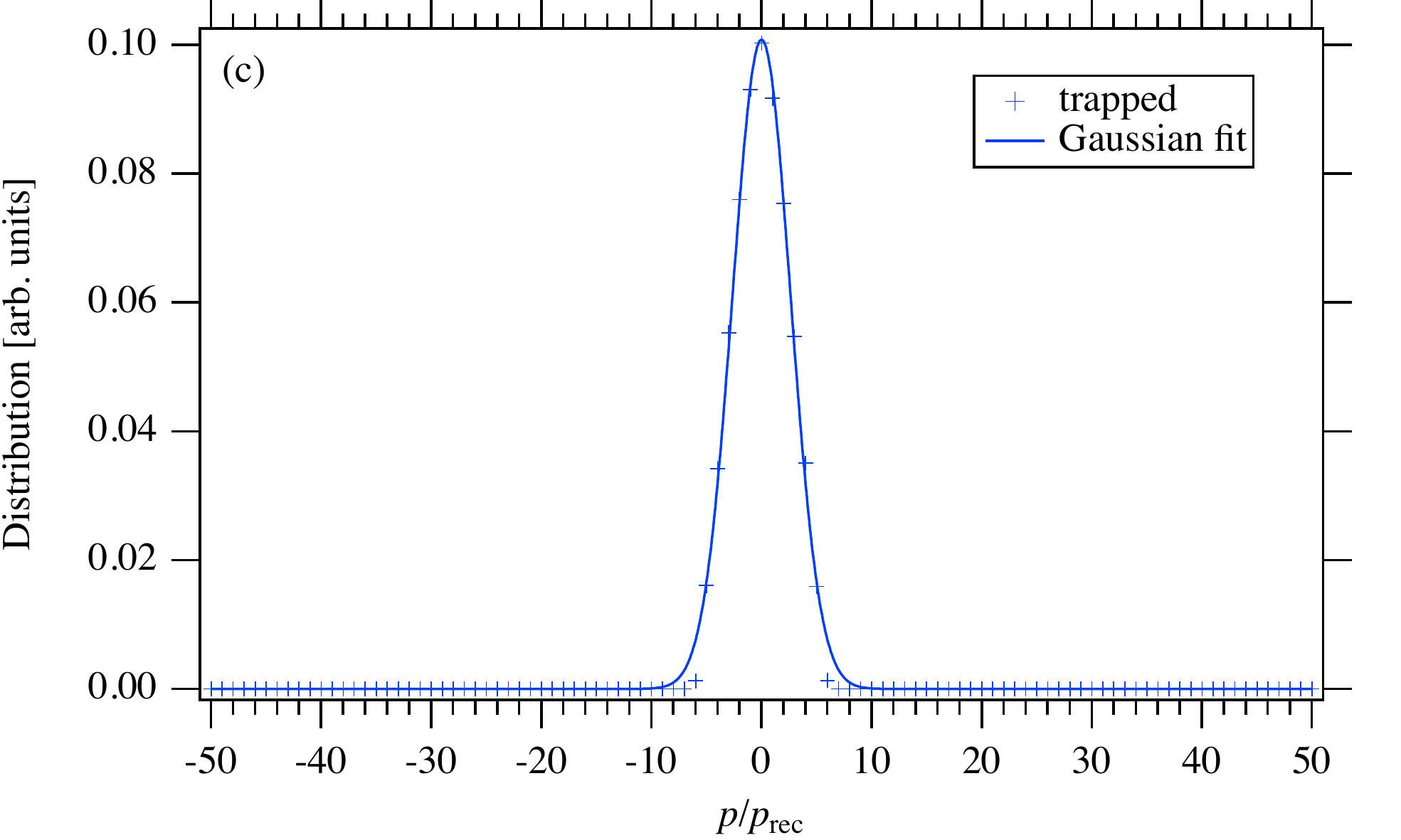}}
	\caption{(Color online) Simulated momentum distribution of atoms cooled on a $1/2 \leftrightarrow 3/2$ transition in an optical lattice, for $\Delta = -10\Gamma$ and $\left|\Delta'\right| = 50 \Er/\hbar$.  (a) Total distribution (full black line), trapped (blue $+$), and untrapped (red $\times$) atoms. (b) Same as (a), but in log scale.  (c) Trapped atoms ($+$) and fit to a Gaussian curve (full line).}
	\label{fig:half_dist}
\end{figure}

We show in Fig.~\ref{fig:half_dist}(a) a typical momentum distribution for the $1/2 \rightarrow 3/2$ transition, for all atoms taken together and for trapped and ``free'' (untrapped) atoms separately (see Sec.~\ref{sec:num_met} for a definition of those terms).  
The central core of the distribution, around $p/\pr = 0$, is mostly made up of trapped atoms, while the tails of the momentum distribution are due to untrapped atoms.  Apart from a small transition region, trapped and untrapped atoms are found at different values of the momentum.  Obviously, no trapped atom can have $p > p_{\mathrm{trap}}$, but we also find that few untrapped atoms have a momentum $p \sim 0$.  This separation in momentum of trapped and untrapped atoms reinforces the conclusions of Refs.~\cite{Jersblad_PRA_2004,Dion_EPL_2005}, where experimental results and quantum simulations of the dynamics of the cooling indicated the presence of ``cold'' and ``hot'' modes in the momentum distribution.

Plotting the same data on a log scale [Fig.~\ref{fig:half_dist}(b)] the distribution of trapped atom appears as an inverted parabola, cutoff at $p_{\mathrm{trap}}/\pr \approx 5.78$ [see Eq.~(\ref{eq:ptrap})].  Indeed, the fit to a Gaussian function is very good, as seen in Fig.~\ref{fig:half_dist}(c).  However, the tails of the total momentum distribution, corresponding to untrapped atoms, do not appear to follow a Gaussian function.  

\begin{figure}
	\centerline{\includegraphics[width=\columnwidth]{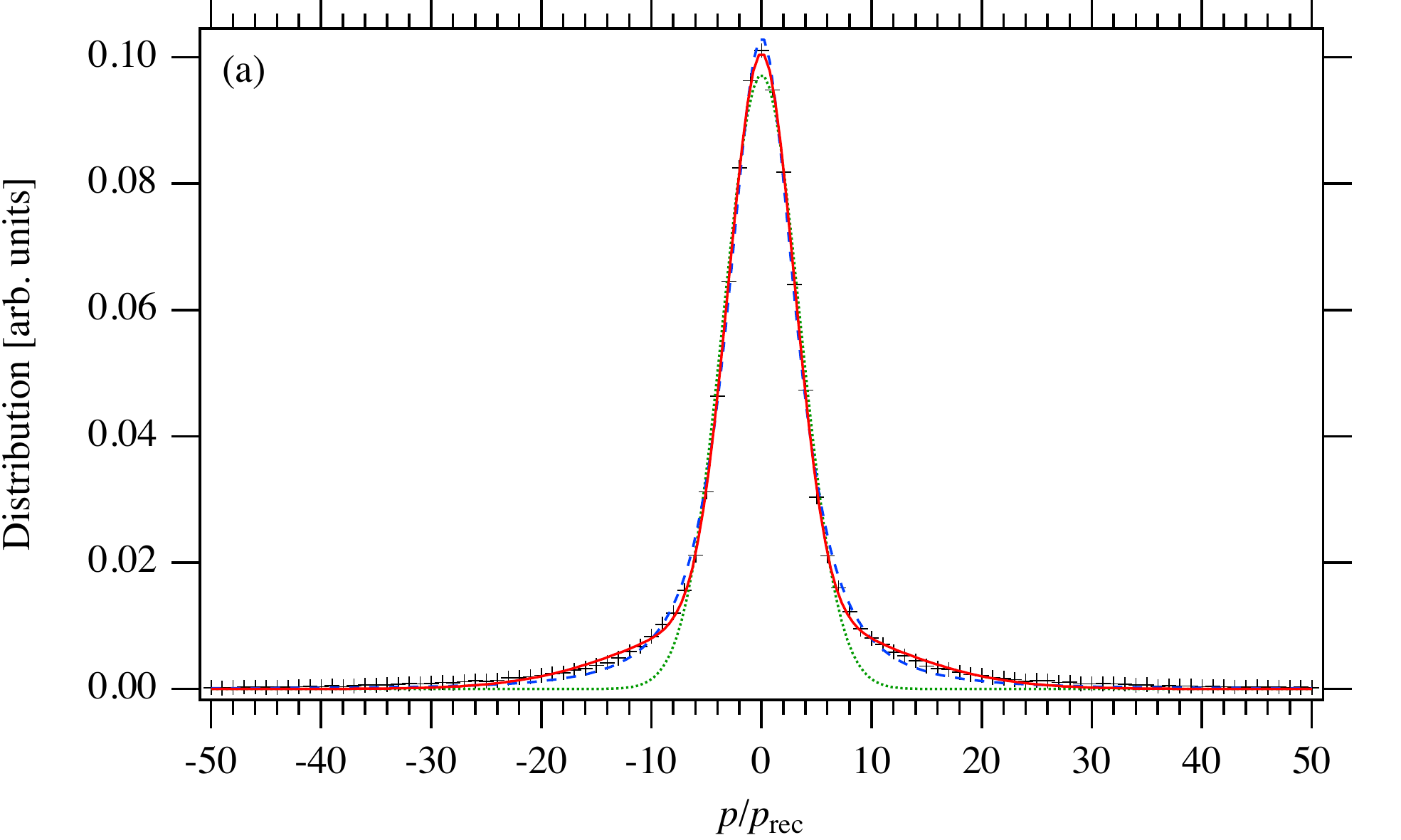}}
    \centerline{\includegraphics[width=\columnwidth]{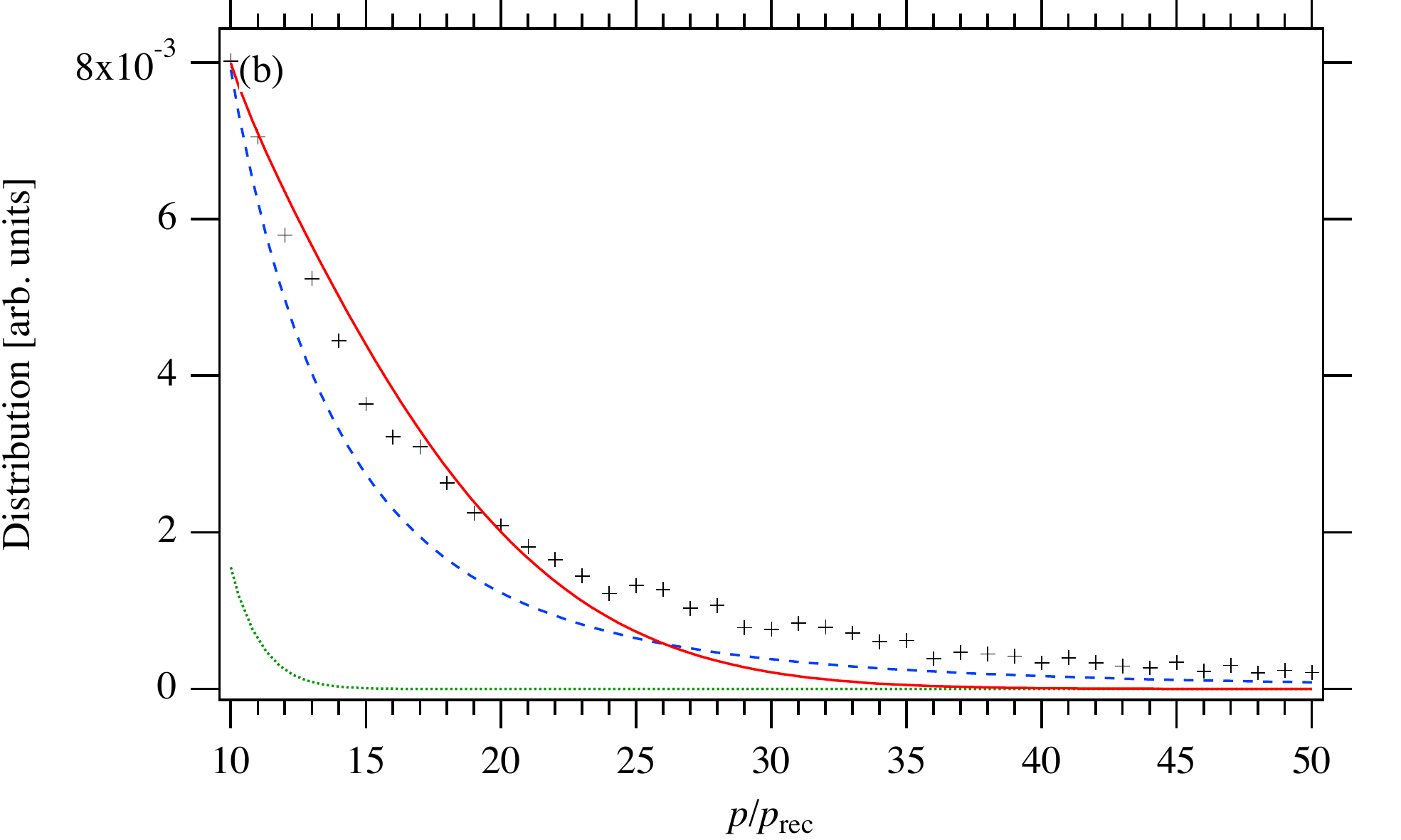}}
	\caption{(Color online) Fits to the \emph{entire} simulated momentum distribution of atoms cooled on a $1/2 \leftrightarrow 3/2$ transition in an optical lattice, for $\Delta = -10\Gamma$ and $\left|\Delta'\right| = 50 \Er/\hbar$.  Panel (b) is a zoom-in of the data shown in panel (a).  In both panels, the simulated data is indicated by crosses, with fits to a single Gaussian (dotted green line), a Tsallis (dashed blue line), and a double Gaussian (full red line) functions. $\chi^2$ values for the fits are \num{7.1e-4}, \num{1.0e-4}, and \num{2.7e-5}, respectively.}
	\label{fig:half_fit}
\end{figure}

To check this further, we fit the \emph{full} data of Fig.~\ref{fig:half_dist} to different functional forms, namely a single Gaussian, a Tsallis function Eq.~(\ref{eq:lutz}), and two Gaussian functions, see Fig.~\ref{fig:half_fit}.
While both the Tsallis function and the double Gaussian reproduce quite well the core of the distribution, this is at the detriment of the tails. 

Fitting the entire dsitribution to a sum of a Gaussian and a power-law function, or to a Gaussian and a Tsallis function, gives an excellent fit. However, this means a function with so much liberty, and so many free parameters, that it is highly questionable if any pertinent conclusion can be drawn from such a fit. Moreover, a power-law function has to be truncated at some point. Instead, we find a fit to a double Gaussian a better indication that the distribution consists of two distict energy modes. In order to test the functional form of the tails of the distribution, a more stringent test is to fit the high-momentum part of the distribution separately. We will address this point in more detail below in Sec.~\ref{sec:powerlaw}.

\begin{figure}
	\centerline{\includegraphics[width=\columnwidth]{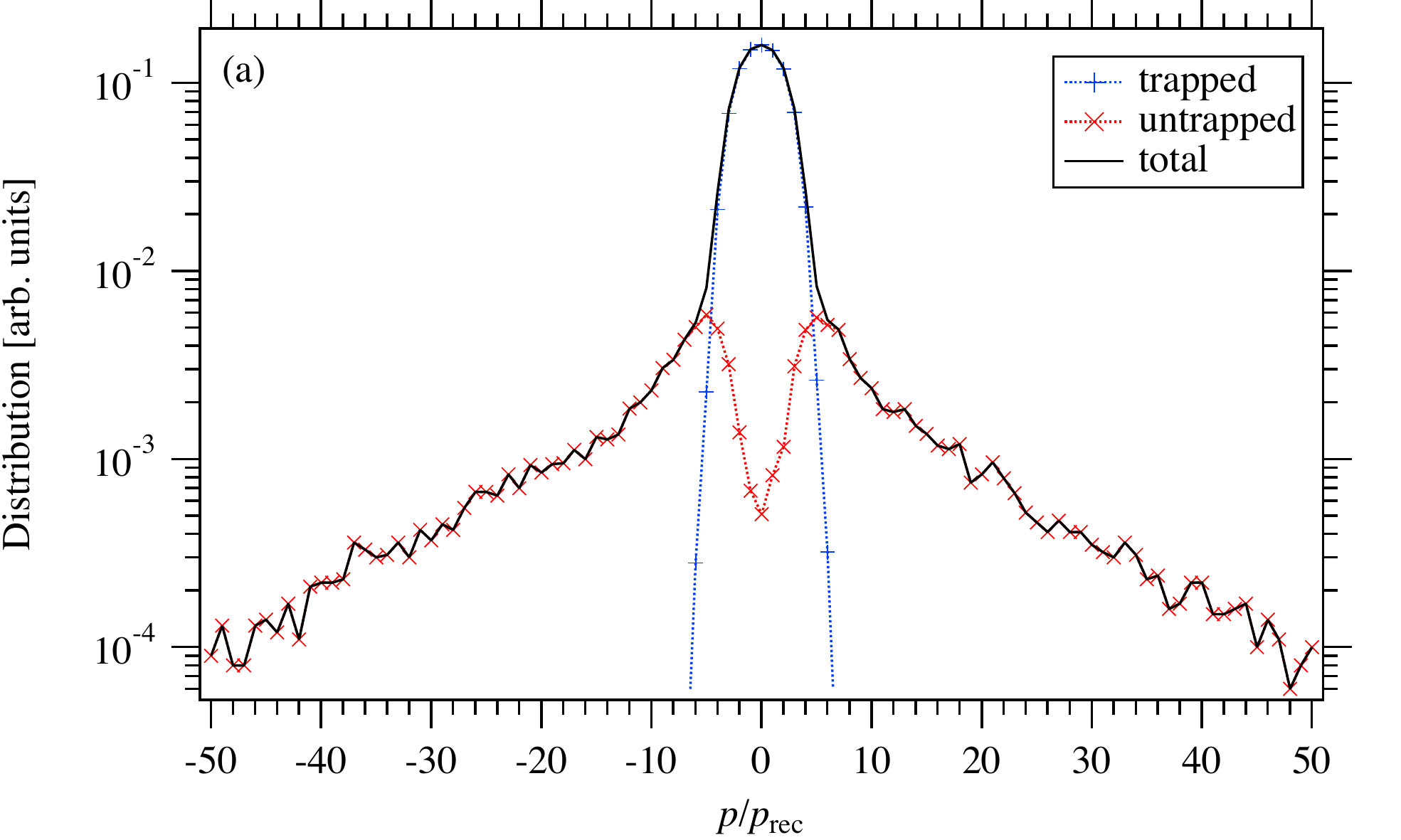}}
    \centerline{\includegraphics[width=\columnwidth]{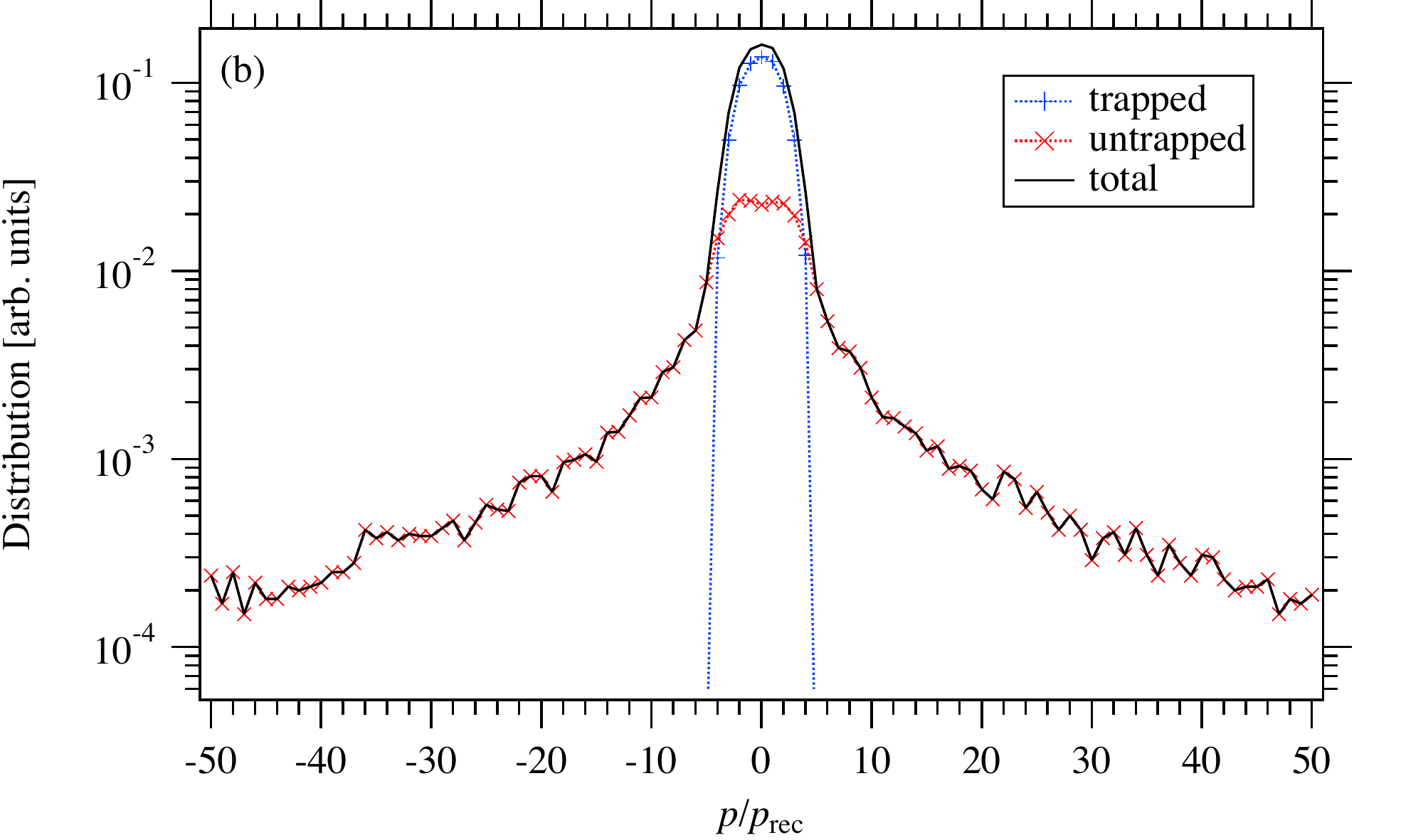}}
	\caption{(Color online) Simulated momentum distribution of atoms cooled on a $4 \leftrightarrow 5$ transition in an optical lattice, for $\Delta = -10\Gamma$ and $\left|\Delta'\right| = 50 \Er/\hbar$, shown on a logarithmic scale. Total distribution (full black line), trapped (blue $+$), and untrapped (red $\times$) atoms; including (a) a single ($F_{\mathrm{e}} = 5$) excited state; (b) two ($F_{\mathrm{e}} = 4,\,5$) excited states.}
	\label{fig:dist_45}
\end{figure}

Similar results are obtained when considering the level structure for the  $F_{\mathrm{g}} = 4 \rightarrow F_{\mathrm{e}} = 5$ transition in cesium, Fig.~\ref{fig:dist_45}, whether including one ($F_{\mathrm{e}} = 5$) or two  ($F_{\mathrm{e}} = 4,\,5$) excited states in the simulation.
Calculating the root-mean-square value of the momentum $p_{\mathrm{rms}}$, we find that the $4 \rightarrow 5$ transition leads to a lower temperature ($p_{\mathrm{rms}}/\pr = 7.52$) compared to the $1/2 \rightarrow 3/2$ transition ($p_{\mathrm{rms}}/\pr = 9.54$), for the same choice of parameters ($\Delta = -10\Gamma$ and $\Delta' = 50 \Er/\hbar$).  There is also stronger trapping for the $4 \rightarrow 5$ transition, with 71.6\% trapped atoms, compared to 65.0\% for the $1/2 \rightarrow 3/2$ transition.  (This may be due to an overestimation in the former case, see Sec.~\ref{sec:num_met}.)

\subsubsection{\label{sec:powerlaw}Power-law tails}

It was shown in Ref.~\cite{Lutz_PRA_2003} that, when neglecting the spatial modulation of the optical lattice and thus the possibility of trapping, the momentum distribution of atoms is in the form of a Tsallis function~\cite{Tsallis_JSP_1988}.  Considering only untrapped atoms, this calculation predicts for a $1/2 \rightarrow 3/2$ transition a tail of the distribution of the form
\begin{equation}
  W(p) = N \left( 1 + \frac{90}{41} \frac{\Delta^2}{\Gamma^2} + \frac{p^2}{p_{\mathrm{c}}^2} \right)^{15 p_{\mathrm{c}} (\Delta/\Gamma) / 41}
  \label{eq:jonsell}
\end{equation}
with $p_{\mathrm{c}} = \hbar \Gamma s_0/(36 \Er)$ and $N$ a scaling constant.  

\begin{figure}
  \centerline{\includegraphics[width=\columnwidth]{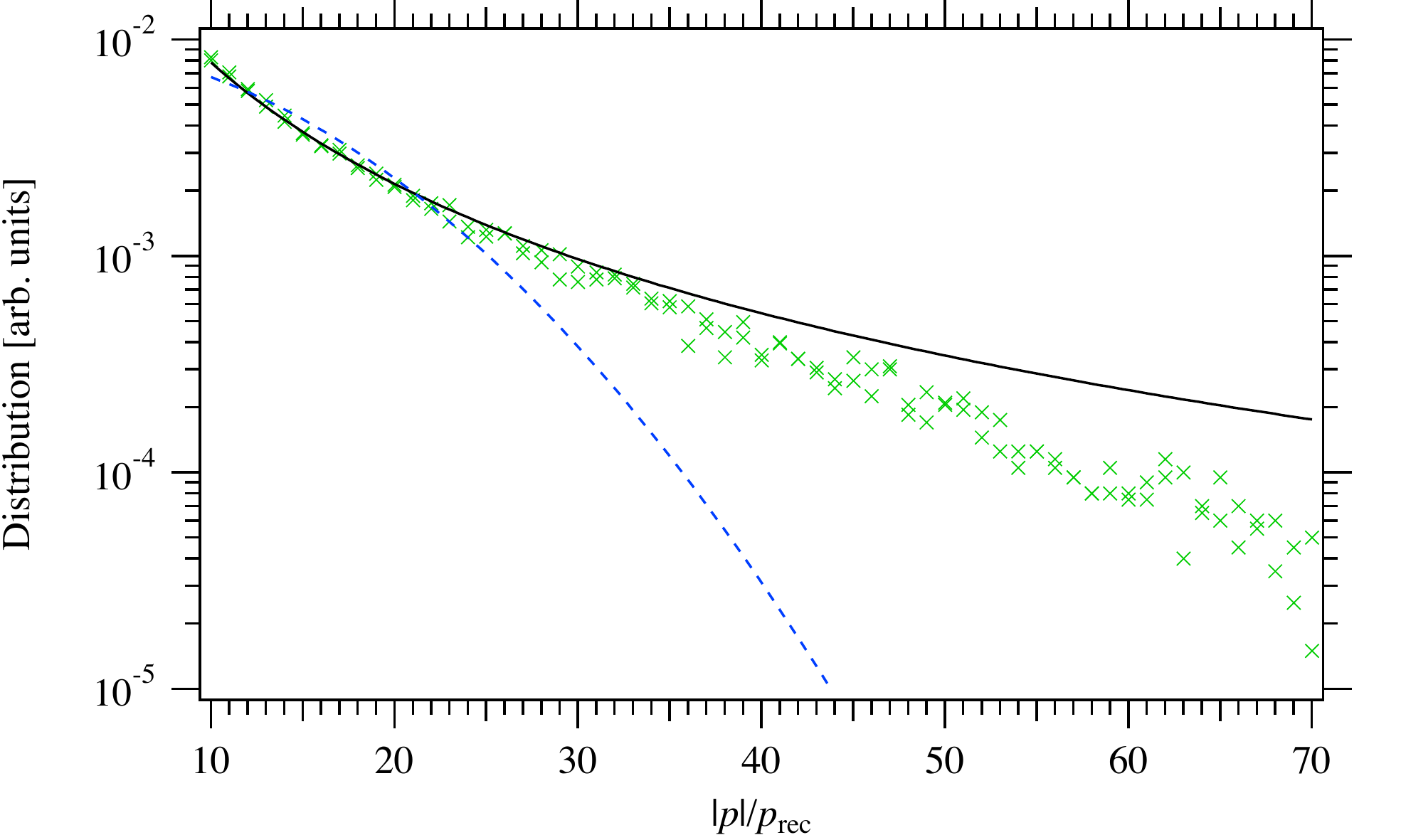}}
  \caption{(Color online) Simulated tails ($\times$) of the momentum distribution shown in Fig.~\ref{fig:half_dist}, for the $1/2 \rightarrow 3/2$ transition ($\Delta = -10\Gamma$, $\left|\Delta'\right| = 50 \Er/\hbar$), with fits to Eq.~(\ref{eq:jonsell}) (full line) and to a Gaussian (dashed line).}
  \label{fig:tail_half}
\end{figure}

We present in Fig.~\ref{fig:tail_half} the data for the tails of the momentum distribution for the $1/2 \leftrightarrow 3/2$ transition  with $\Delta = -10\Gamma$ and $\left|\Delta'\right| = 50 \Er/\hbar$ (same as in Fig.~\ref{fig:half_dist}).
We have combined here the data points for both negative and positive momenta.  We have fitted the data for the tail separately, selecting the part of the distribution with $\left| p \right| / \pr \geq 10$, with Eq.~(\ref{eq:jonsell}), using $N$ as the \emph{only} free parameter, as $p_{\mathrm{c}}$ can be expressed in terms of the simulation parameters $\Delta$ and $\Delta'$.  The result (full line in Fig.~\ref{fig:tail_half}) agrees very well with the simulated data, especially for smaller values of the momentum (the statistics get worse as the momentum increases, as very few atoms reach high momenta in the simulation).  For comparison, we have also fitted the tail to a Gaussian function (dashed line in Fig.~\ref{fig:tail_half}), and the result clearly shows that the momentum of untrapped atoms does not follow a normal distribution.  This would indicate that the system is non-ergodic~\cite{Lutz_PRL_2004}.

While the theory presented in Sec.~\ref{sec:NG_tails} appears to work well for untrapped atoms, the bimodal nature of the distribution argues against the use of a single function to describe the \emph{entire} momentum distribution, as it appears that a good fit of the core of the distribution results in an incorrect description of the tail, see Fig.~\ref{fig:half_fit}. It is clear that a double Gaussian function cannot capture all properties of the distribution, but nevertheless it does capture the bimodality, and it does provide a better fit than the Tsallis function, when the \emph{entire} population is included in the fit.


\section{\label{sec:exp}Experiments}
In order to further investigate the velocity distribution, we perform an experiment with a three-dimensional optical lattice. The experimental set-up has been described in detail elsewhere (\eg, in Ref.~\cite{Hagman_JAP_2009}), and therefore the present description is kept brief.

A cold sample of atoms is prepared by stopping a thermal beam of cesium, followed by the loading of the atoms in a magneto-optical trap (MOT). The atoms are then progressively cooled by going through stages of a low-intensity MOT, a low-intensity optical molasses, and eventually the atoms are loaded in a three-dimensional dissipative optical lattice. This traps the upper hyperfine structure state of the ground configuration, 6s~$^2$S$_{1/2}$, $F_\mathrm{g}=4$. 

\begin{figure}
	\centerline{\includegraphics[width=.7\columnwidth]{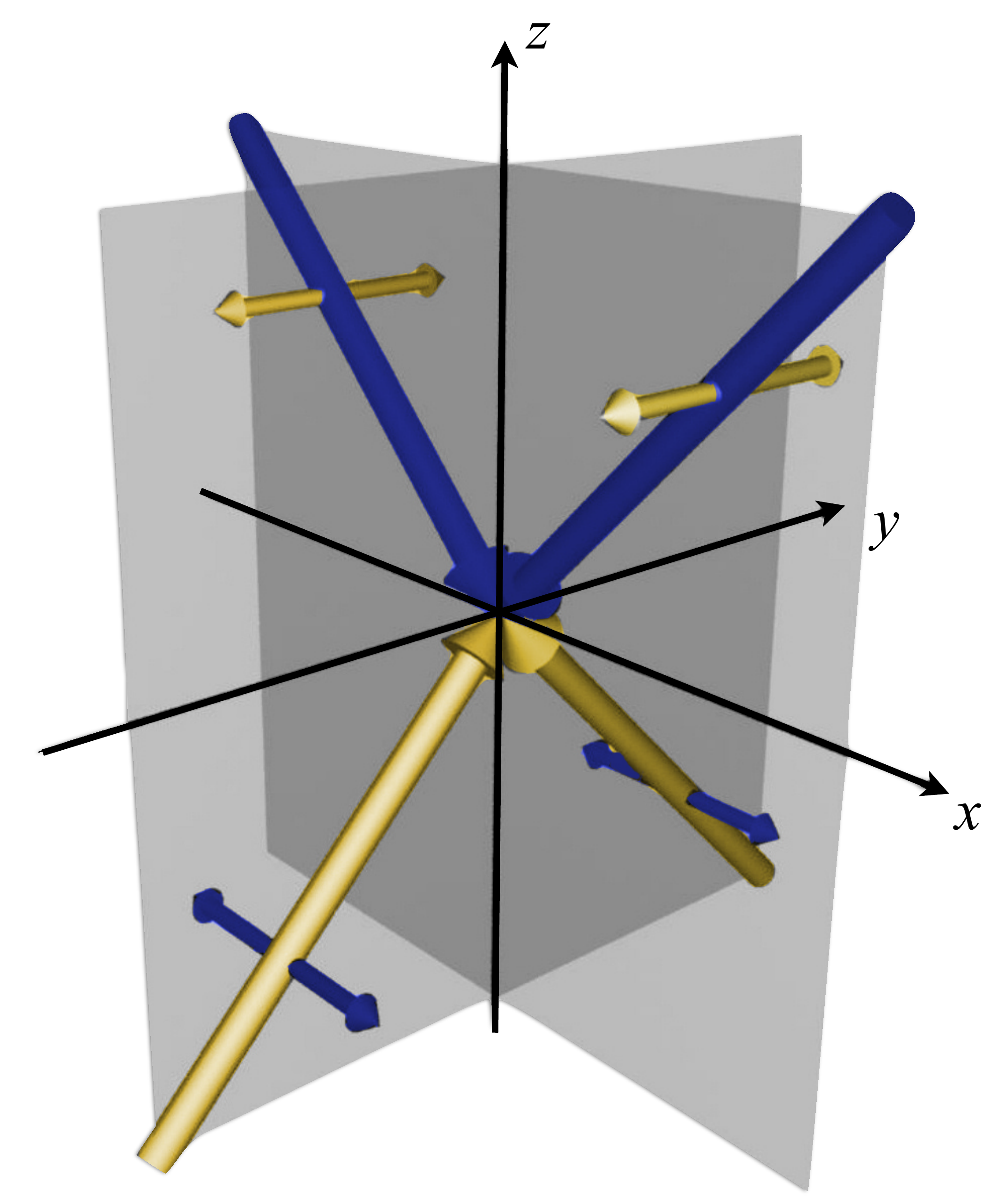}}
	\caption{(Color online) Configuration of the three-dimensional optical lattice. A red-detuned beam is split into four beams. These are aligned, and their polarizations are chosen, as shown in the figure. This provides a three-dimensional generalization of the one-dimensional lin$\perp$lin laser cooling configuration.}
	\label{fig_latticefig}
\end{figure}

The optical lattice configuration is shown in Fig.~\ref{fig_latticefig}. Four laser beams with identical detunings and intensities make an angle of $\pi/4$ with the principal axis ($\hat{z}$), with the latter being parallel to the vertical axis. Two beams are in the $xz$-plane and are polarized along $\hat{y}$, whereas the other two are polarized along $\hat{x}$ and propagate in the $yz$-plane. The lasers are typically detuned by $\Delta=-25\Gamma$ from the resonance $F_\mathrm{g}=4 \leftrightarrow F_\mathrm{e}=5$ in the D2 line of Cs (see, \eg, \cite{Steck_Cs_2010}), at $\lambda\approx852$ nm, but this may be varied. That is, the light-atom interaction is in a regime where the kinetics of the atoms are strongly influenced by incoherent scattering, which includes both laser cooling and momentum diffusion.

\begin{figure}
\centerline{\includegraphics[width=.75\columnwidth]{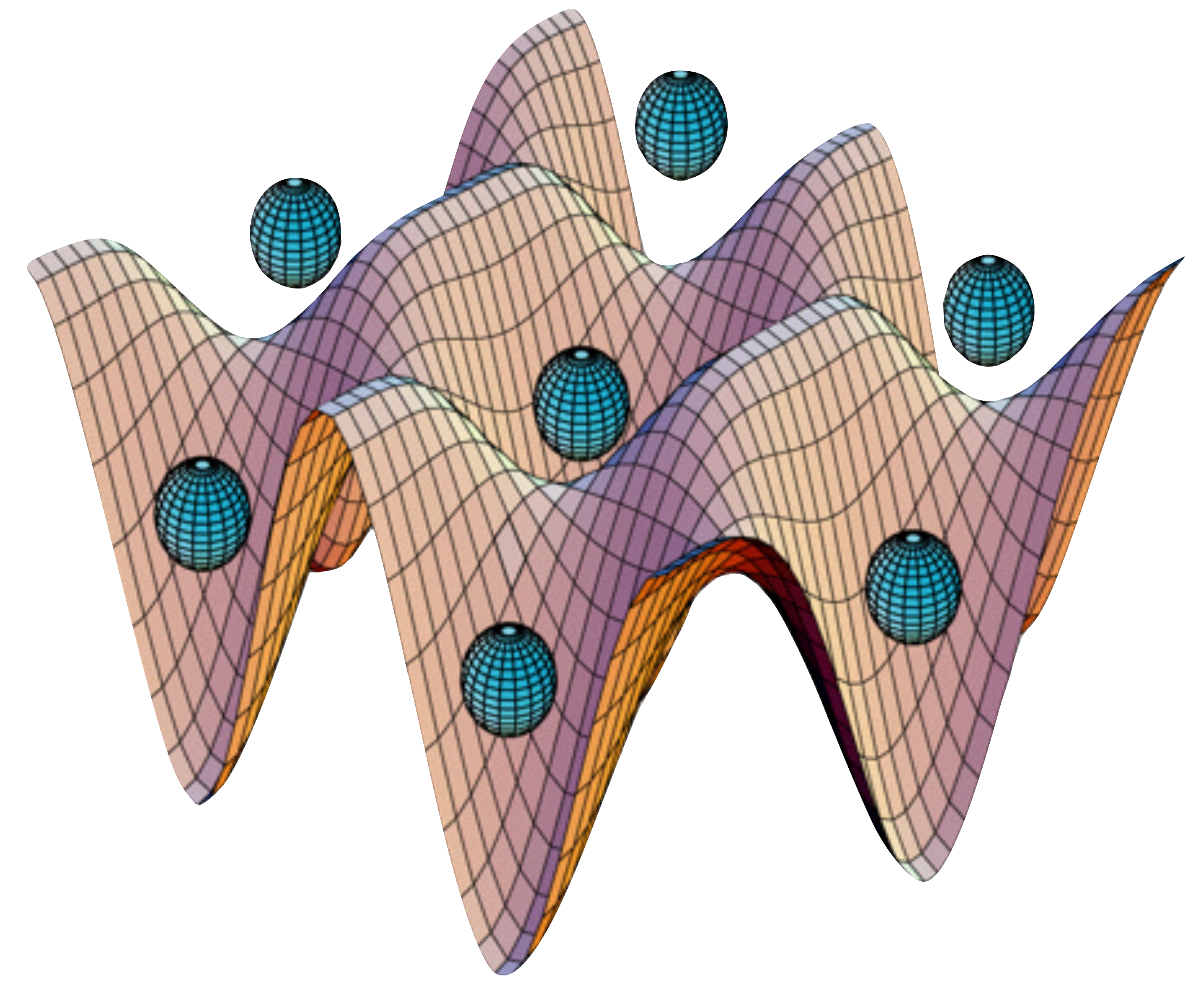}}
	\caption{(Color online) Two-dimensional projection of the calculated lowest adiabatic potential (experimentally verified in Ref.~\cite{Ellmann_PRL_2003}) of the optical lattice. Along the $\hat{z}$-direction (in the figure, this correspond to the diagonal from top left to bottom right), the modulation is purely sinusoidal.}
	\label{fig_eggcrate}
\end{figure}

The resulting optical lattice potential is illustrated in Fig.~\ref{fig_eggcrate} (the lowest adiabatic potential is shown). Along the vertical $\hat{z}$-axis, we have a sinusoidal potential, and a cooling configuration that closely corresponds to the one-dimensional lin$\perp$lin configuration, and thus also to the model used in Sec.~\ref{sec:sc_sim}. The potential depth scales proportionally to $I/\Delta$ (see eq.~\ref{eq:depth}), with $I$ the laser irradiance, and is thus tunable.

The velocity distributions are observed by using the time-of-flight method~\cite{Hagman_JAP_2009}, where atoms are released from the optical lattice and are allowed to expand under free fall. The expansion is then measured by a laser probe. In our experiments, we obtain a signal-to-noise ratio in the measured velocity distributions better than 1:1000 (in a single shot), and a velocity resolution better than \SI{10}{nK}. The maximum repetition rate is of the order of one hertz, and thus, good statistics can easily be obtained.

\subsection{\label{sec:exp_res}Experimental results}
\begin{figure}
\centerline{\includegraphics[width=\columnwidth]{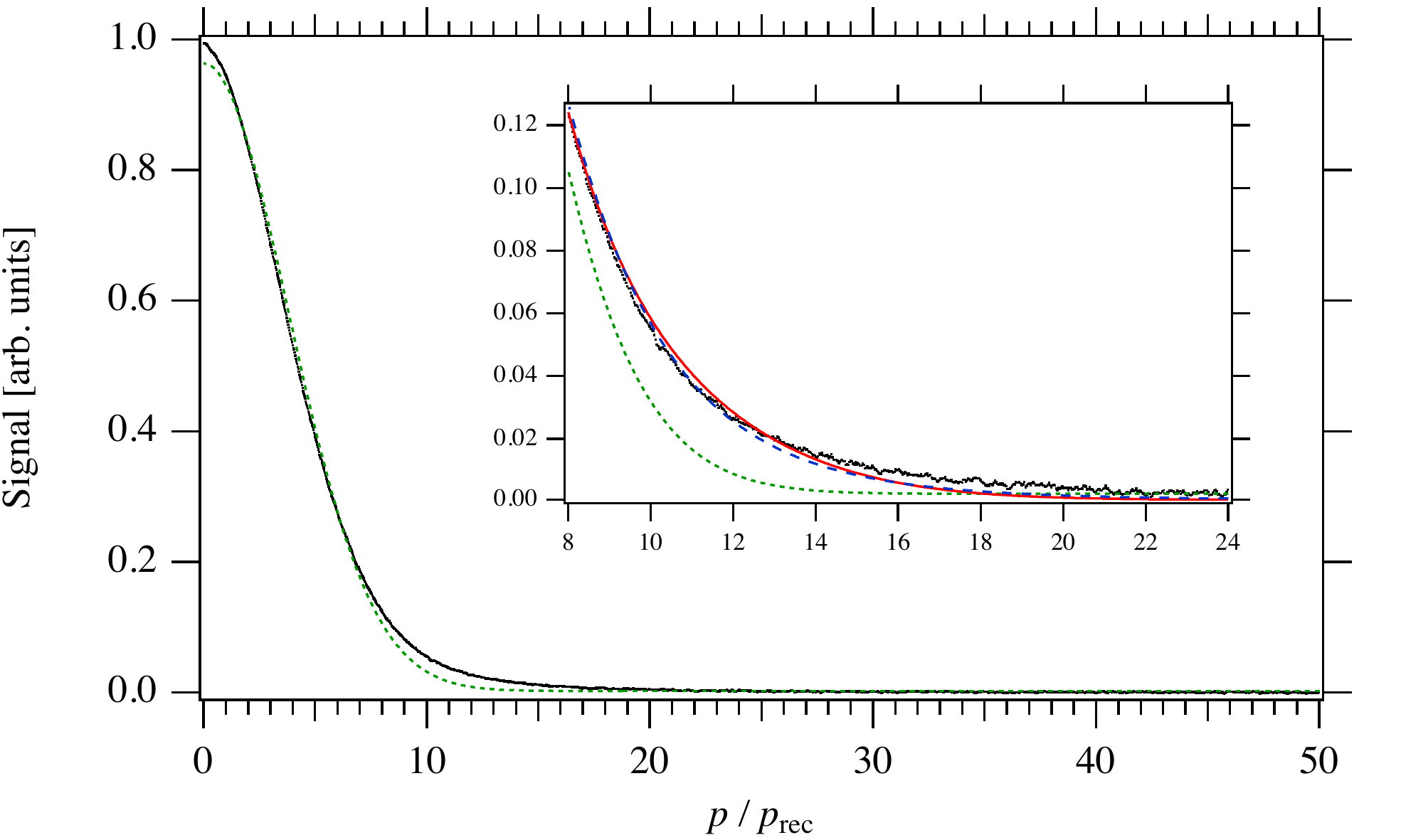}}
	\caption{(Color online) Velocity distribution recorded for a shallow optical lattice ($U = 106 E_\mathrm{rec}$), close to the critical potential depth (average of two measurements). In the full figure, the dotted green line is a Gaussian function fitted to the data. Fits to a Tsallis function, or to a double Gaussian, are visually indistinguishable from the experimental data at this scale, and are therefore not included. In the inset is a zoom-in of the wing of the distribution. The dotted green line is still the single Gaussian, whereas the dashed blue line is a fit to a Tsallis function, and the full red line one to a double Gaussian.}
	\label{fig_experiment}
\end{figure}

The velocity distribution has been recorded for a range of potential depths. In Fig.~\ref{fig_experiment} we show a result, for the case of very low intensity, and hence very shallow light-shift potentials ($U/E_\mathrm{rec}=106$, obtained for a laser power of $P= \SI{0.13}{mW}$ per beam). As has been previously shown (see, \eg, \cite{Gatzke_PRA_1997,Jersblad_PRA_2000,Jersblad_PRA_2004}) a fit to a Gaussian function of the velocity distribution gives an estimate of the kinetic temperature of the sample, and above a certain critical potential depth --- of the order of 10--$100 \Er$ --- the temperature scales linearly with potential depth ($I/\Delta$). 

For deep potentials, Gaussian fits to the velocity distributions are excellent. Close to the critical point (as in Fig.~\ref{fig_experiment}), such fits are still fairly good, but there is a systematic underestimation of the wings of the distribution. There is a high-velocity tail that cannot be mimicked by a Maxwell-Boltzmann distribution. 

\begin{figure}
\centerline{\includegraphics[width=\columnwidth]{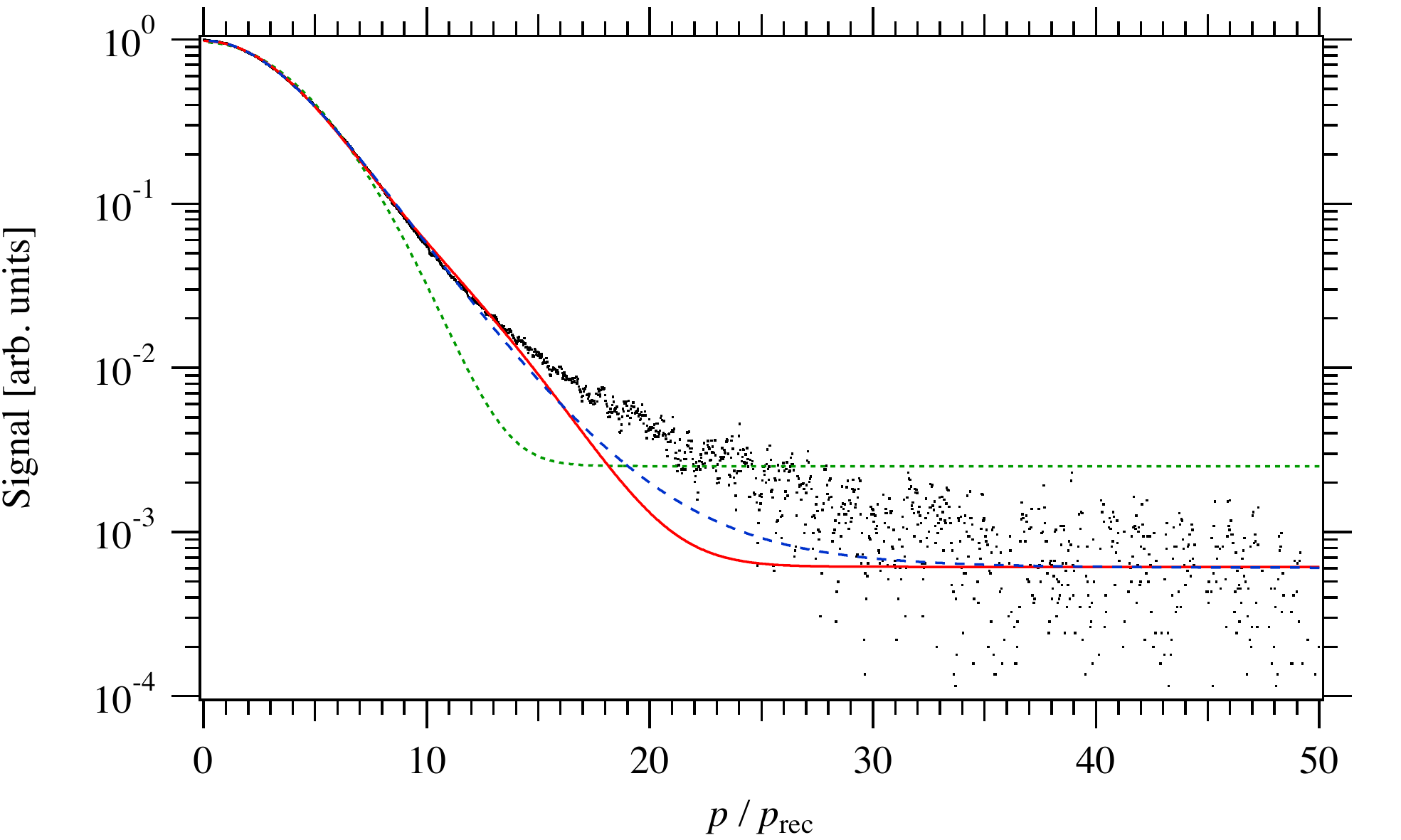}}
	\caption{(Color online) The same data as in Fig.~\ref{fig_experiment}, but shown in a logarithmic scale. The mismatches of the fits, as quantified by their $\chi^2$-values are: $\chi^2=0.16$ for a single Gaussian (dotted green line), $\chi^2=0.0068$ for a Tsallis function (dashed blue line), and $\chi^2=0.011$ for a double Gaussian (full red line). For all fits, the $y$-intercept is a fitting parameter, which explains why the fits flatten out. The influence of this on the least-squares fit is negligible, since it concerns data three decades smaller than the center of the distribution.}
	\label{fig_exp_log}
\end{figure}

In Fig.~\ref{fig_experiment} --- and also in Fig.~\ref{fig_exp_log}, which is the same data plotted in a logarithmic scale --- we show fits of various functions to the data. 
A Tsallis function gives a good fit, and so does a double Gaussian. However, when the Tsallis function is fitted to the entire distribution, it gives systematicalle different fit results than it does when only the high-velocity tail is fitted. The double Gaussian is just the simplest bimodal model, and it is noteworthy that it still provides as good a fit as does the Tsallis function.

The results supports the assumption that a significant part of the atomic
population is localized in optical lattice sites. For deep lattices, this proportion is close to 100\%. Closer to the critical point, a gradually larger proportion of the atoms will, on average, be untrapped, and therefore analyzing the entire population in terms of one single distribution function will not give a fully pertinent description. This is totally consistent with the numerical results in Sec.~\ref{sec:sc_sim} and with experimental results reported by others in, \eg, Refs.~\cite{Westbrook_PRL_1990,Hamilton_PRA_2014}.


\section{\label{sec:discussion}Discussions}

\begin{figure}
\centerline{\includegraphics[width=\columnwidth]{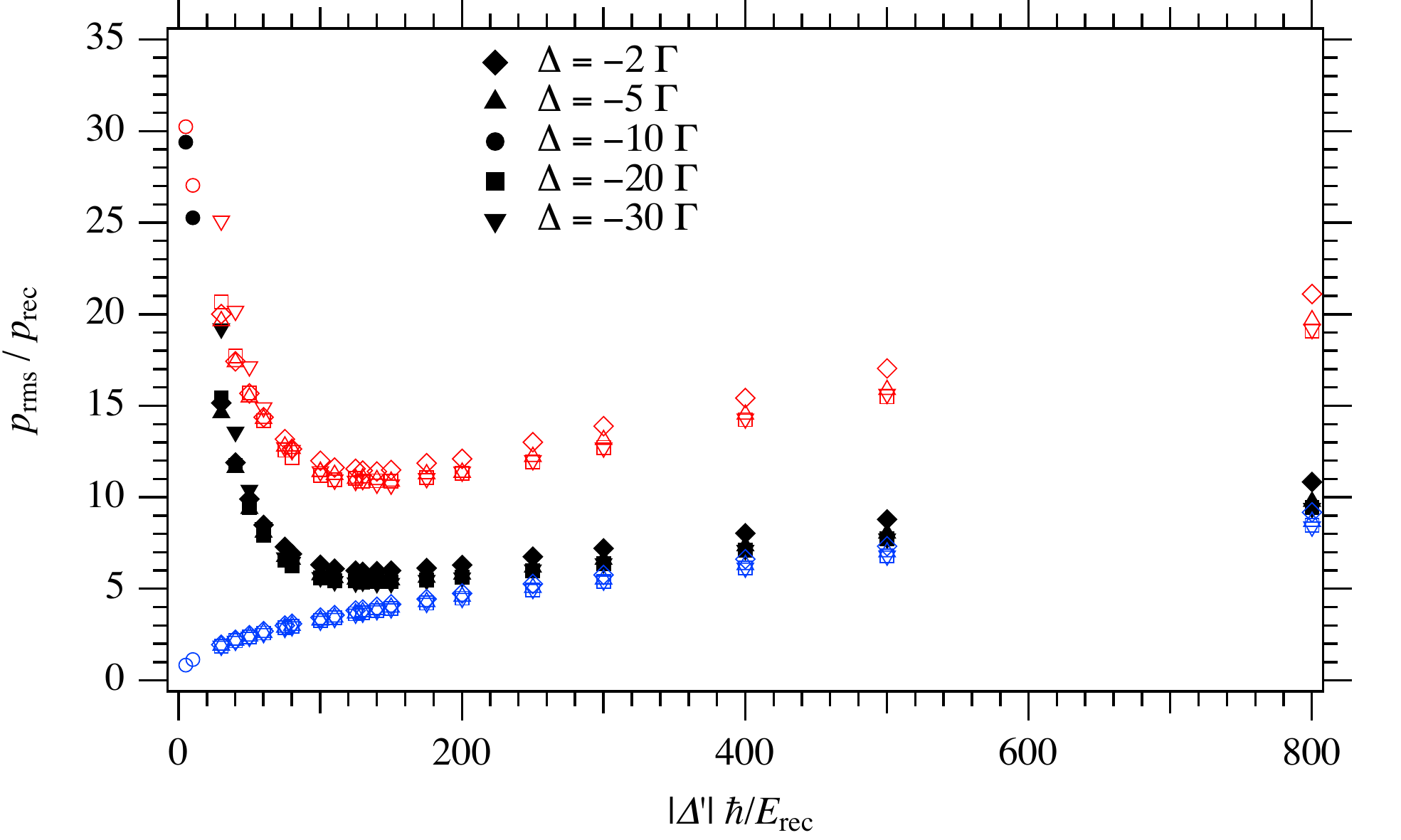}}
\centerline{\includegraphics[width=\columnwidth]{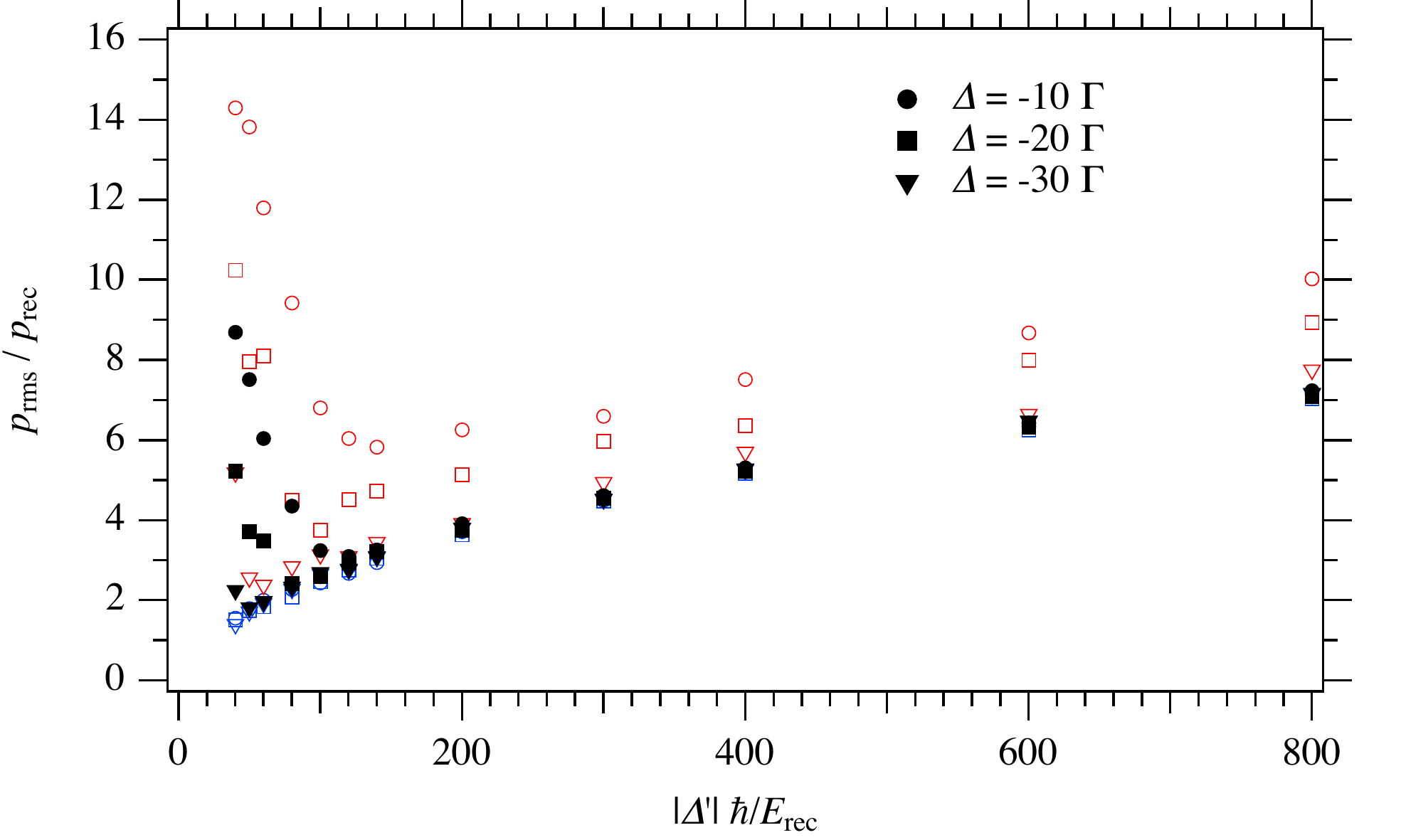}}
  \caption{(Color online) Root-mean-square momentum for all atoms (filled symbols); and for trapped (blue open symbols, lower points) and untrapped (red open symbols, upper points) only.  (a) $1/2 \leftrightarrow 3/2$ transition (simulations done for 50000 atoms); (b) $4 \leftrightarrow 5$ transition (simulations done for 5000 atoms).}
  \label{fig:prms}
\end{figure}

\begin{figure}
  \centerline{\includegraphics[width=\columnwidth]{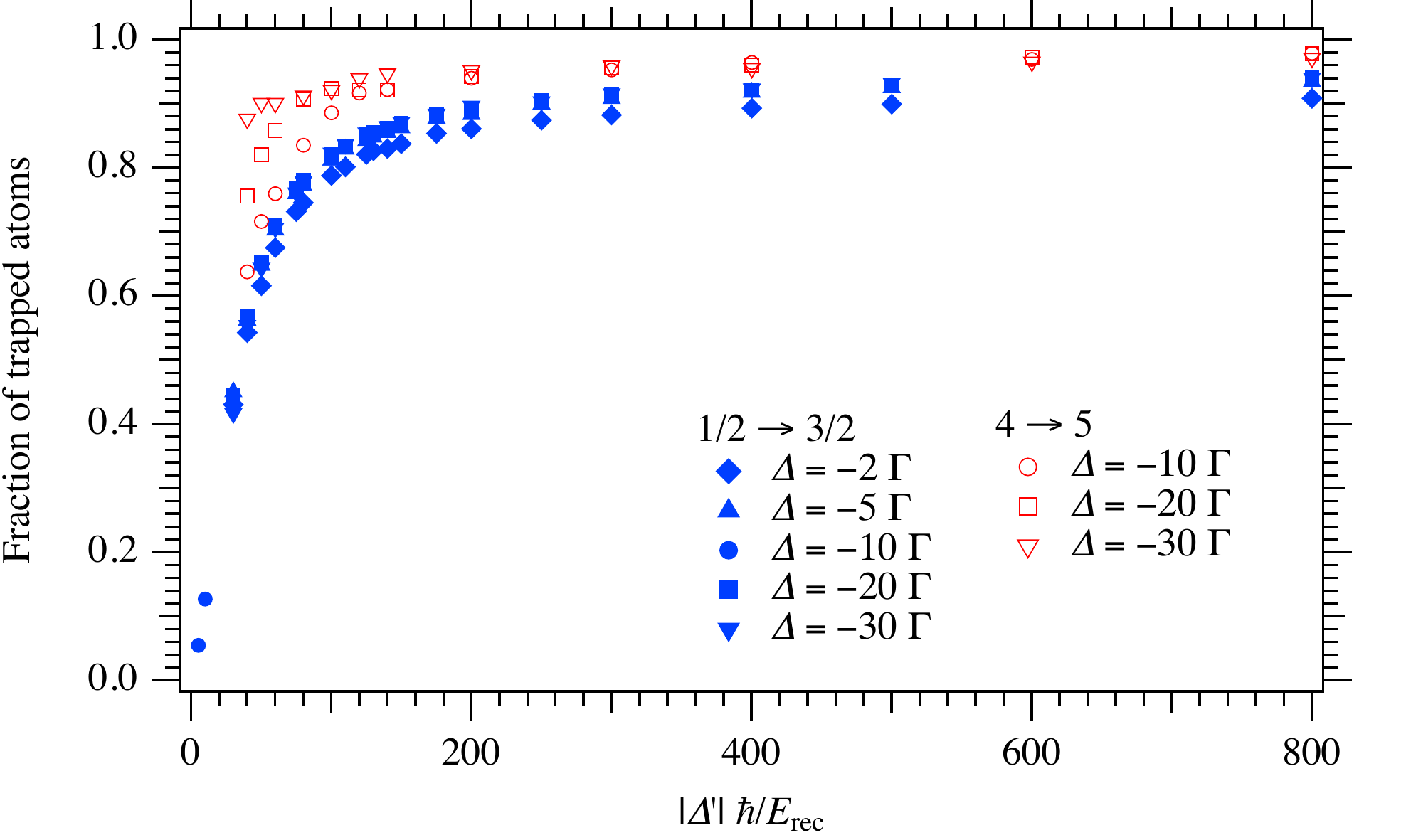}}
  \caption{(Color online) Fraction of trapped atoms, for $1/2 \leftrightarrow 3/2$ (filled symbols) and $4 \leftrightarrow 5$ (open symbols) transitions (simulations done for 50000 and 5000 atoms, respectively).}
  \label{fig:trapped}
\end{figure}

The observation that the atoms are found in two modes, trapped and untrapped, allows us to revisit a striking feature of laser cooling by optical lattices, the \emph{d\'ecrochage} mentioned in Sec.~\ref{sec:NG_tails}.  This phenomenon can be seen in Fig.~\ref{fig:prms}, where the root-mean-square momentum $p_{\mathrm{rms}}$ obtained from the numerical simulations is plotted as a function of $\Delta'$, which is directly proportional to the potential depth.
It now appears that only untrapped atoms are responsible for the \emph{d\'ecrochage} phenomenon observed in an experiment.  While the value of $p_{\mathrm{rms}}$ of the trapped atoms varies monotonously with $\Delta'$, the momentum of untrapped atoms increases as the potential depth goes below the threshold of \emph{d\'ecrochage}.  This effect is magnified by the fact that the proportion of trapped atoms is significantly reduced for shallow potentials, as seen in Fig.~\ref{fig:trapped}. We notice the greater trapping of atoms with a higher degeneracy of the ground state.  We also point out that a significant portion of the atoms remain trapped even past \emph{d\'ecrochage}, and that it is only for extremely shallow potentials that the majority of atoms are not trapped. In an experiment, the average of \emph{all} atoms will be measured, and even though few atoms are untrapped, the very high momentum of these will give rise to the observed ``unhooking'' (departure) of the recorded data from the linear intensity dependence.

Two additional remarks on Fig.~\ref{fig:prms} are in order.  First, for a two-level system ($1/2 \leftrightarrow 3/2$ transition) there is only a slight influence of the detuning $\Delta$ on the values of $p_{\mathrm{rms}}$ obtained, with higher values obtained for smaller detunings, at a given value of $\Delta'$.  Within the range $\Delta = -10 \Gamma$ to $-30 \Gamma$, the values do not differ by more than the statistical noise of the simulations.  However, a difference becomes clear at $\Delta = -5 \Gamma$, while the values at $-2 \Gamma$ stand out even at the scale of the figures presented here.  We find a similar result when we consider the $4 \leftrightarrow 5$ transition \emph{without} the presence of the $F_{\mathrm{e}} = 4$ state (see also~\cite{Sanchez-Palencia_EPJD_2002}).  This is not the case for the three-level $4 \leftrightarrow 5$ transition (\ie, with the $F_{\mathrm{e}} = 4$ state included), Fig.~\ref{fig:prms}(b), where the position of \emph{d\'ecrochage} is clearly influenced by the detuning, as previously noted in Ref.~\cite{Svensson_EPJD_2008}.  This is in agreement with experimental results for both cesium~\cite{Jersblad_PRA_2000} and rubidium~\cite{Carminati_EPJD_2001}, where it was found that \emph{d\'ecrochage} appears at constant laser irradiance (meaning constant $s_0$ in our model, so that the potential depth becomes directly proportional to the detuning $\Delta$).  This dependence on the detuning is also reflected in the fraction of trapped atoms, see Fig.~\ref{fig:trapped}.

For full disclosure all data used for the figures in this article are published in Ref.~\cite{Dion_figshare_2016}, in order to enable further analysis by others.

\section{\label{sec:conclusion}Conclusions}

Our results, both experimental and numerical, strongly support the assumption that the velocity distribution of atoms trapped in a shallow, dissipative optical lattice is bimodal. We have found nothing that supports an hypothesis that at some potential depth, near \emph{d\'ecrochage}, there is a sudden transition between a localized regime and a jumping one. Rather, our data supports the theory that atoms that are constantly exposed to both laser cooling and heating in a dissipative optical lattice go through periods of being trapped as well as of being untrapped. At any given moment, the entire population will consist of these two modes. 

For deep optical lattices, the untrapped portion will be very small. Closer to \emph{d\'ecrochage} (\ie, for decreasing potential depth) a gradually larger subset of the ensemble will have enough energy to move over more than one lattice site. For these shallow potentials, a fit to a single distribution function of the entire population cannot be adequately applied, and any theory that applies spatial averaging of the atomic density, over several lattice sites, will fail to account for the significant portion of the atoms that remain trapped.

For the untrapped atoms, a power-law distribution such as Eq.~(\ref{eq:lutz}) gives a good fit to numerical data. Also for the experimental data, the high-velocity tail of the distribution clearly deviates from a simple Gaussian, but it is more difficult to prove a power-law distribution. For a detailed experimental study of a sample displaying non Boltzmann-Gibbs statistics, as suggested in, \eg, Refs.~\cite{Lutz_PRL_2004,Holz_EPL_2015,Dechant_PRL_2015}, a different physical system than a pure dissipative optical lattice would be needed. This could for example be a weak Sisyphus cooling configuration superimposed on an external potential. In that case, trapping could be avoided, and spatial averaging can be applied in the analysis. Examples of this is a cooling inside an ion trap, as in Ref.~\cite{Katori_PRL_1997}, or weak cooling in an external optical trap, as in Ref.~\cite{Sagi_PRL_2012}.

\begin{acknowledgments}
We thank R.\ Kaiser and L.\ Sanchez-Palencia for stimulating discussions. This research was funded in part by the Swedish Research Council (VR).  The computations were performed on resources provided by the Swedish National Infrastructure for Computing (SNIC) at the National Supercomputer Centre (Link\"oping University).
\end{acknowledgments}

\bibliography{biblio}

\end{document}